# Three pathways of cell transformation of lymphoid cell: a slow, a rapid, and an accelerated

Jicun Wang-Michelitsch[1]*, Thomas M Michelitsch[2]

[1] Independent researcher

[2] Sorbonne Université, Institut Jean le Rond d'Alembert, CNRS UMR 7190 Paris, France

## Abstract

Lymphoid leukemia (LL) and lymphoma are two groups of neoplasms developed from lymphoid cells (LCs). To understand why different forms of LL/lymphoma occur at different ages, we analyzed the effects of different types of DNA changes on a LC and the cellular characteristics of LCs. Point DNA mutations (PDMs) and chromosome changes (CCs) are the two major types of DNA changes. CCs have three subtypes by their effects on a LC: great-effect CCs (GECCs), mild-effect CCs (MECCs), and intermediate-effect CCs (IECCs). PDMs and MECCs are mostly mild thus can accumulate in cells. Some of the PDMs/MECCs contribute to cell transformation. A GECC affects one or more genes and can alone drive cell transformation. An IECC affects one or more genes and participates in cell transformation. Due to some cellular characteristics, a LC may have higher survivability from DNA changes and require obtaining fewer cancerous properties for transformation than a tissue cell. Hence, a LC can be more rapidly transformed by a CC. On this basis, we hypothesize that a LC may have three pathways on transformation: a slow, a rapid, and an accelerated. Slow pathway is driven by accumulation of PDMs/MECCs through many generations of cells. Rapid pathway is driven by a GECC in "one step", namely in one generation of cell. Accelerated pathway is driven by accumulation of PDMs/MECCs/IECC(s) through a few generations of cells. Cell transformations of a LC via different pathways occur at different ages. A transformation via slow pathway occurs mainly in adults. A transformation via rapid pathway occurs at any age and has no increasing incidence with age. A transformation via accelerated pathway occurs also at any age but has increasing incidence with age. In conclusion, a LC may have three pathways on cell transformation, and the occurring age of LL/lymphoma may be determined by the transforming pathway of a LC.

## Keywords

**This paper has the following structure:**

**I.  Introduction**

**II.  Chromosome changes (CCs) are more frequent in pediatric lymphoid leukemias (LLs) and lymphomas than in adult cases**

 2.1 Two types of DNA changes: point DNA mutations (PDMs) and CCs
 2.2 CCs are more frequent in pediatric LLs/lymphomas than in adult cases

**III.  Different types of DNA changes have different effects on a lymphoid cell (LC)**

3.1 Effect of a PDM on a LC
3.2 Three groups of CCs by their effects on a LC
   3.2.1  Great-effect CCs (GECCs)
   3.2.2  Mild-effect CCs (MECCs)
   3.2.3  Intermediate-effect CCs (IECCs)

**IV.  A LC may have higher survivability from DNA changes than a tissue cell**

4.1 Distinct cellular characteristics of a LC from a tissue cell
4.2 Higher survivability of a LC from DNA changes than a tissue cell
4.3 A differentiating LC may have higher tolerance to DNA changes than a non-differentiating LC

**V.  A LC may require obtaining fewer cancerous properties for transformation than a tissue cell**

**VI.  Transformation of a LC can be driven by accumulation of PDMs, MECCs, and IECCs or by a GECC**

6.1 All LCs can be transformed by accumulation of PDMs and MECCs
6.2 A differentiating LC may have a risk to be transformed by a GECC in "one step"
6.3 An IECC may accelerate the cell transformation driven by accumulation of PDMs and MECCs

**VII. Our hypothesis: a LC may have three pathways on cell transformation**

7.1 Slow pathway: by accumulation of PDMs and MECCs through many generations of cells
7.2 Rapid pathway: by a GECC in "one step" in one generation of cell
7.3 Accelerated pathway: by accumulation of PDMs, MECCs, and IECC(s) through a few generations of cells

**VIII.  Three grades of transformation of a LC: low, high, and intermediate**

**IX.  Summary**

**X.  Conclusions**

## I. Introduction

Lymphoma and lymphoid leukemia (LL) are two groups of blood cancers developed from lymphoid cells (LCs). Different forms of LL/lymphoma tend to develop at different ages (Mori, 2016). For example, acute lymphoblastic leukemia (ALL) and Burkitt lymphoma (BL) have both a peak incidence in young children, and Hodgkin's lymphoma (HL) has the highest incidence at age 20s (Kelly, 2015). Chronic lymphocytic leukemia (CLL) and mantle cell lymphoma (MCL) occur mainly in old people (Lehmann, 2016). Diffuse large B-cell lymphoma (DLBCL) occurs mostly in adults but also in children (Swerdlow, 2016). Development of childhood LL/lymphoma indicates that a LC can be transformed more rapidly. Forms of cytogenetic abnormalities and gene mutations are identified in pediatric LLs/lymphomas (Merker, 2007; Blombery, 2015; Bogusz, 2016). However, it is unknown what a type of DNA change is associated with the rapider transformation of LC in childhood blood cancer.

We have discussed recently the potential sources of cell injuries of LCs and the mechanism of generation and accumulation of DNA changes in LCs (Wang-Michelitsch, 2018a). In the present paper, we will discuss why a LC can be transformed more rapidly. We will analyze the cellular characteristics of a LC distinct from a tissue cell and the effects of different types of DNA changes on a LC. "LC" is a general name for all the cells in lymphoid lineage at any developing stage. Thus, LCs include all the B-cells and T-cells in marrow, thymus, bloodstream, lymph nodes (LNs), and lymphoid tissues (LTs). The term "lymphocyte" refers to uniquely the naïve lymphocyte. We aim to show by our discussion that, a LC may have three pathways on transformation by different types of DNA changes: a slow, a rapid, and an accelerated. A transformation via rapid pathway is made by "one step", namely in one generation of cell. A transformation via slow pathway is made by many steps (through many generations of cells). A transformation via accelerated pathway is made by a few steps (through a few generations of cells).

We use the following abbreviations in this paper:

| | |
|---|---|
| AM: acquired mobility | IECC: Intermediate-effect chromosomal change |
| AML: acute myeloid leukemia | LC: lymphoid cell |
| ALCL: anaplastic large cell lymphoma | LCCI: loss of cell-contact inhibition |
| ALL: acute lymphoblastic leukemia | LN: lymph node |
| ATLL: adult T-cell lymphoma/leukemia | LT: lymphoid tissue |
| BL: Burkitt lymphoma | MALTL: mucosa-associated lymphoid tissue lymphoma |
| CC: chromosomal changes | |
| CML: chronic myeloid leukemia | MCL: mantle cell lymphoma |
| CLL: chronic lymphocytic leukemia | MDS: myelodysplastic syndrome |
| DLBCL: diffuse large B-cell lymphoma | MECC: mild-effect chromosomal change |
| EBV: Epstein-Barr virus | NCC: numerical chromosome change |
| FL: follicular lymphoma | NHL: non-Hodgkin's lymphoma |
| GECC: great-effect chromosomal change | PDM: point DNA mutation |
| HIV: human immunodeficiency virus | PMM: production of matrix metalloproteinases |
| HL: Hodgkin's lymphoma | SCC: structural chromosome change |
| HLTV-1: human T cell lymphotropic virus type 1 | SIM: stimulator-independent mitosis |
| HSC: hematopoietic stem cell | T-LBL: T-lymphoblastic lymphoma/leukemia |

## II. Chromosome changes (CCs)are more frequent in pediatric lymphoid leukemias (LLs) and lymphomas than in adult cases

DNA changes are the trigger for cell transformation in tumor development. Point DNA mutation (PDM) and chromosome change (CC) are the two major types of DNA changes. PDMs and CCs are both found in leukemia cells and lymphoma cells. A PDM can affect at most "one" gene, thus it is called also gene mutation and molecular mutation. CCs can be seen under microscope by specific staining, thus a CC is called also cytogenetic abnormality and abnormal karyotype. A karyotype that has no visible and detectable CC is regarded as normal karyotype. Clinic investigations show that CCs are more frequent in pediatric LLs and lymphomas than in adult cases (Martinez-Climent, 1997).

### 2.1 Two types of DNA changes: point DNA mutations (PDMs) and CCs

A PDM is a DNA change on one or two bases, such as alteration, deletion, or insertion of a base in a DNA. PDMs (gene mutations) are the most common type of DNA changes in tumor cells, but they may exist also in normal cells. Studies showed that generation of PDM is a result of Misrepair of DNA rather than an error of DNA synthesis. In our view, PDMs are generated and accumulate as a consequence of repeated cell injuries and repeated cell proliferation (Wang-Michelitsch, 2015).

A CC can be numerical CC (NCC) or structural CC (SCC). A NCC exhibits as loss (-) or gain (+) of one or more chromosomes of a cell. For example, the term (-7) refers to loss of a chromosome 7 of a cell. Hyperdiploid (or hypodiploid) refers to the karyotype that a cell has slightly more (or less) chromosomes than normal. A cell that has hyperdiploid may have a deformed nucleus. Aneuploid refers to the karyotype that the chromosomes in a cell are not in diploid. Aneuploid is the most common form of DNA change in pediatric ALL (Harrison, 2013). Gain or loss of an arm of chromosome, such as (+1q), (+7q), del (6q), del (13q), and del (17p), are often seen in later stage of LL/ lymphoma (Lones, 2006). A NCC is generated rather as a consequence of dysfunction of cell division promoted by damage.

A SCC exhibits as rearrangement of part of a chromosome, such as translocation (t), deletion (del), gain (+), inversion (inv), and amplification (amp) of portion of a chromosome. Chromosome translocation exhibits that a segment of chromosome (chromosome-A) attaches by mistake to another chromosome (chromosome-B) or to another location of chromosome-A. Chromosome amplification exhibits that a segment of chromosome has additional copies. For example, intra-chromosomal amplification of chromosome 21 is observed in some ALLs (Harrison, 2016). Chromosome inversion exhibits that part of a chromosome is inversed in direction. For example, inv (16) has high recurrence in acute myeloid leukemia (AML) (Estey, 2013). Among all forms of SCCs, chromosome translocation has the highest frequency in LLs/lymphomas. Cryptic SCCs may exist in a cell which has normal karyotype. The karyotype that has more than three forms of CCs is called complex karyotype. In HL, the H/R-S cells have often complex karyotypes (Weniger, 2006). A SCC is generated often as a result of Misrepair of DNA on multiple DNA breaks (Wang-Michelitsch, 2018a).

## 2.2  CCs are more frequent in pediatric LLs/lymphomas than in adult cases

Cancers occur rarely in children. However, leukemia and lymphoma are two exceptions. Some forms of LL/lymphoma including ALL and BL occur mainly in children. Pediatric forms of blood cancers differ from adult forms not only by pathology but also by DNA changes in tumor cells. In general, CCs are more frequent in pediatric LLs/lymphomas than in adult cases (Table 1). For example, in ALL, which occurs mostly in children, 72% of cases have CCs (Al-Bahar, 2010; Braoudaki, 2012). The most frequent CCs in ALL are aneuploid (in 50% of cases) and t(12;21) (25%). In CLL, which occurs mainly in old people, CCs are rare (Amaya-Chanaga, 2016). In lymphoma, BL, T-lymphoblastic lymphoma/leukemia (T-LBL), and ALK-positive anaplastic large cell lymphoma (ALK$^+$-ALCL) are three forms of lymphomas that have higher incidences in children and adolescents than in adults. The most frequent forms of DNA changes in these lymphomas are CCs, for example, t(8;14) in BL, t(1;14) in T-LBL, and t(2;5) in ALK$^+$-ALCL (Smith, 2010; Bonzheim, 2015; Fielding, 2012).

**Table 1. Chromosome changes are more frequent in pediatric LLs/lymphomas than in adult cases**

| Pediatric LL/lymphoma | Recurrent DNA changes | | Adult LL/lymphoma | Recurrent DNA changes |
|---|---|---|---|---|
| ALL | Aneuploid or t(12;21) | | CLL | Gene mutations |
| T-LBL | t(1;14) or t(10;14) | | FL | Gene mutations and t(14;18) |
| BL | t(8;14) | | MCL | t(11;14) and *Sox11* mutation |
| Pediatric DLBCL | MYC-translocation in 30% of cases | | Adult DLBCL | MYC-translocation in 10% of cases |
| ALK$^+$-ALCL | t(2;5) | | ATLL | Gene mutations |

Follicular lymphoma (FL), MCL, and adult T-cell lymphoma/leukemia (ATLL) are adult forms of lymphomas. Gene mutations are the main type of DNA changes in these lymphomas. Exceptionally, two forms of CCs including t(14;18) and t(11;14) have high recurrences in follicular lymphoma (FL) and MCL, respectively. However, t(14;18) and t(11;14) are detectable also in normal individuals who have no lymphoma. This indicates that these two forms of CCs have only mild effects on a LC. DLBCL occurs mostly in adults but also in children. However, the cell-of-origin and the driver DNA changes in pediatric DLBCLs are different from that in adult cases. Adult DLBCL can arise from a germinal center B-cell (GCB) or a plasmablast (called also activated B-cell (ABC)). Differently, pediatric DLBCLs arise mostly from GCB cell (immunoblast). MYC-translocation is found in 10% of adult GCB-DLBCLs, but in 30% of pediatric cases (Table 1). Adult DLBCLs have higher frequency of gene mutations, such as *C-rel* amplification, *EZH2* mutation, and *PTEN* deletion (Oschlies, 2006; Nedomova, 2013; Dobashi, 2016). Taken together, CCs have much higher frequency in pediatric forms of LL/lymphoma than in adult forms. This indicates that CCs play a critical role in the occurrence of pediatric LL/lymphoma. CCs should be responsible for the rapider cell transformation of a LC.

### III. Different types of DNA changes have different effects on a lymphoid cell (LC)

In a cell, different DNAs (chromosomes) contain different genes which code different proteins. However, only a small part of a DNA (in one chromosome) is used for "genes". Namely, most parts of a DNA are non-coding for proteins. Thus, the DNA changes generated in different DNAs and in different parts of a DNA will have different effects on a cell. But one thing is sure: the bigger a DNA change is, the more genes will it affect. Thus a CC has greater effect than a PDM on a cell.

### 3.1 Effect of a PDM on a LC

A PDM affects only one or two bases of a DNA, thus it is often silent or mild to a cell. If it occurs to a non-coding part of a DNA, it may have no direct effect on the cell. If it occurs to the non-determining part of a three-base genetic code (for an amino-acid), it may not affect the expression of a protein. Therefore, most PDMs generated in a cell are silent. However, if a PDM occurs in a coding region of a DNA, it may result in alteration of expression of a protein. The final result can be one of the followings: **A.** failure of expression of the protein, **B.** altered level of expression of the protein, or **C.** production of mutant and un-functional proteins. Failure of expression of a protein can be a result of mutation in the promoter sequence of a gene or deletion/addition of a base (causing frame-shift mutation) in the coding part of a gene. Altered level of protein expression is often due to a mutation in the enhancing sequence of a gene. Production of mutant proteins is often a result of alteration of a base in the coding part of a gene. Consequently, if the affected gene is essential for cell survival, failure or insufficiency of gene expression or production of non-functional proteins by a PDM may lead to cell death. In this case, the PDM can be fatal. However, if the affected gene is not essential for cell survival but expressed only by inductions, the PDM may not necessarily cause cell death. In this case, the PMD is mild.

Therefore, most PDMs are silent or mild, thus they can "survive" and accumulate in cells. By accumulation, some mild PDMs may exhibit their accumulated or emerged effects on a cell. If some of the PDMs make a cell proliferate independently, the cell is transformed. The PDMs that contribute to cell transformation are driver mutations, and others are passenger mutations. Some PDMs may alter neither cell phenotype nor cell functions, but affect genome stability and enhance generations of other DNA changes. These PDMs may also contribute to cell transformation.

### 3.2 Three groups of CCs by their effects on a LC

In general, a CC has greater effect on a cell than a PDM, because it may affect multiple genes. However, different forms of CCs may have different effects on a cell. The effect of a CC on a cell depends on the size and the location of a CC in a chromosome. Some CCs are fatal to cells, while others may have only mild effect. To understand the roles of different types of CCs in cell transformation of a LC, in this paper, we classify CCs into three groups by their effects on a LC: great-effect CCs (GECCs), mild-effect CCs (MECCs), and intermediate-effect CCs (IECCs) (Box 1).

### 3.2.1 Great-effect CCs (GECCs)

GECC is a type of CC that affects one or more genes and can alone cause cell transformation (Box 1). By affecting one or more critical genes in a cell, a GECC will cause cell death in most cases (in these cases, the GECCs cannot be observed). However, in a rare case, for certain types of cells, a GECC may be not fatal, but rather transforming. Through altering the expressions of one or more oncogenes and/or tumor-suppressing genes, a GECC can alone drive cell transformation. In this case, the cell is transformed by a GECC in "one step". We will discuss why a LC can be transformed rapidly by a GECC in next parts of this paper. Probably, those forms of CCs that are associated with poor prognoses (high risk) of diseases in LLs/lymphomas are GECCs. For example, three forms of CCs, including the aneuploid in ALL, the t(2;5) in ALK$^+$-ALCL, and the t(8;14) in BL, might be GECCs. Since a GECC is determinant for cell transformation, it can be used as a prognostic factor. A GECC is also an ideal blocking target in targeted cancer treatment.

**Box 1. Different types of DNA changes have different effects on a lymphoid cell**

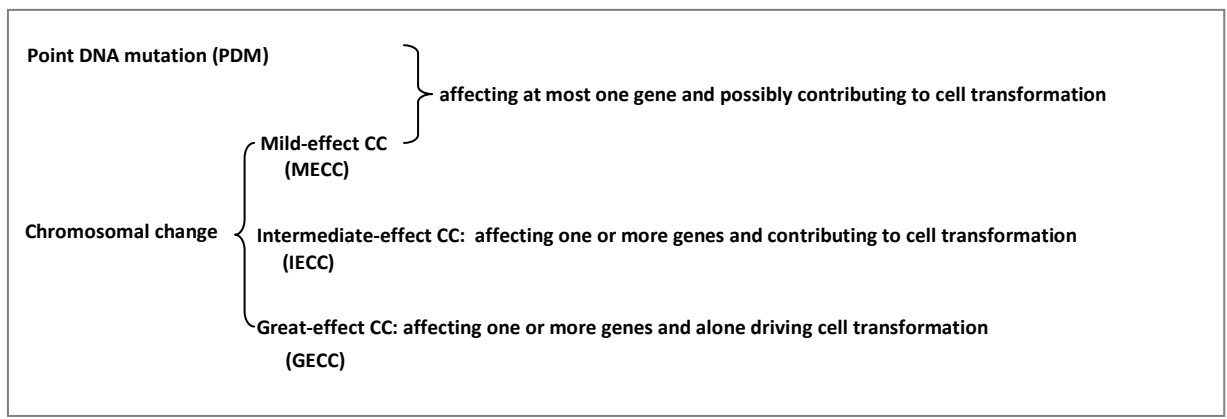

### 3.2.2 Mild-effect CCs (MECCs)

MECC is a type of CC that has silent or mild effect on a cell. Thus on effect, a MECC is similar to a PDM (Box 1). Since a NCC can affect multiple genes, MECCs are mainly tiny SCCs. MECCs do not cause cell death, thus they can accumulate in cells like PDMs. Some MECCs may contribute to cell transformation. For example, the t(14;18) in FL and the t(11;14) in mantle cell lymphoma (MCL) may be two forms of MECCs (Kishimoto, 2014). The DNA translocation in t(14;18) leads to generation of fusion gene of *IGH-BCL2* and increased expression of BCL2 protein. However, t(14;18) is found also in normal people who have no lymphoma (Rabkin, 2008). t(11;14) occurs to 90% of MCLs, but it is also found in 60% of CD20+ multiple myeloma. These data suggest that t(14;18) and t(11;14) are early events of cell transformations respectively in FL and MCL. MECCs might represent the cytogenetic changes that are related to favorable (very low risk) prognoses of cancers. For example, the

t(15;17) in AML and the del (13q14) in CLL may be also MECCs (Estey, 2013; Sagatys, 2012).

### 3.2.3 Intermediate-effect CCs (IECCs)

IECC is a type of CC that affects one or more genes and contributes to cell transformation (Box 1). Differently from a GECC, an IECC cannot alone cause cell death or cell transformation. An IECC can be generated in a cell at any time as a consequence of cell injuries and DNA injuries. However, a cell that has already other DNA changes will have increased risk of generation of IECC, because some PDMs/MECCs can affect the structural stability of chromosomes in a cell. Namely, the more PDMs/MECCs a cell has, the more risk has a cell on generation of an IECC. It is known that the somatic cells of an organism will accumulate more and more mutations with age of the organism. Thus, IECCs have increasing occurrence in cells with age of our body. IECCs may also accumulate in cells, but this probability is low. Since generation of an IECC is often related to early DNA changes in the cell, the cell transformation driven by IECC is a co-effect of the IECC and other PDMs/MECCs. With stronger effect than a PDM/MECC, an IECC can accelerate cell transformation. An IECC drives cell transformation possibly by affecting an oncogene or a tumor suppressing gene.

A good example of IECC is the Ph translocation (t(9,22)) in chronic myeloid leukemia (CML) and in Philadelphia chromosome positive ALL (Ph$^+$-ALL) (Neviani, 2014). Ph translocation drives cell transformation by generating fusion gene of *BCR-ABL1* and activating permanently the tyrosine kinase of ABL1. CML occurs rarely in young children, and it has increasing incidence with age. Ph$^+$-ALLs represent 20% of adult ALLs but only 3% of pediatric ALLs. These data suggest that Ph translocation has increasing occurrence in LCs with age. Other examples of IECCs are: the CRLF2 rearrangement in Ph-like ALL, the MYC-translocations in DLBCL, and the (+8) in AML (Estey, 2013). These three diseases, including Ph-like ALL, DLBCL, and AML, are similar to CML on occurring age, since they all occur in older children and adults. IECCs might represent the cytogenetic abnormalities that are associated with intermediate-risk or high-risk prognoses of cancers. Clinic data have shown that, these above-mentioned CCs, including Ph translocation, MYC-translocations, CRLF2 rearrangement, and (+8), are poor prognostic factors in blood cancers. An IECC is a driver DNA change, thus it can be chosen to be blocking-target in targeted cancer treatment.

## IV.  A LC may have higher survivability from DNA changes than a tissue cell

Cytogenetic abnormalities are often seen in leukemia cells and lymphoma cells but rarely seen in cancers and sarcomas developed from tissue cells. This phenomenon indicates that a LC may have higher survivability from DNA changes than a tissue cell. It is known that EBV-infection is associated with two forms of cancers: BL and nasopharyngeal cancer (Raab-Traub, 2015). However, nasopharyngeal cancer occurs mainly in adults but BL occurs mainly in children. CCs are rare in nasopharyngeal cancer cells but frequent in BL cells. This difference suggests that BL may develop by a more rapid cell transformation of LC which is driven by CCs. A LC is more tolerant to DNA changes than a tissue cell possibly because it is a type of blood cell.

## 4.1 Distinct cellular characteristics of a LC from a tissue cell

As a type of blood cell and a type of immune cell, a LC is different from a tissue cell by some cellular properties. Firstly, a LC is anchor-independent for cell survival, but a tissue cell is anchor-dependent. A normal tissue cell needs to survive in tissue environment. If a tissue cell fails to anchor to neighbor cells, it will undergo apoptosis. This is called anoikis. However, for anchoring to neighbor cells or to extracellular matrixes (ECMs), a tissue cell needs to express a pattern of cell adhesion molecules, including cadherins, selectins, and integrins. Most of these molecules are expressed in a tissue cell constitutively. Sufficient expression of these proteins is essential for the survival of a tissue cell. Namely, failure of expression of one of these proteins by a DNA change will lead to death of a tissue cell.

Differently, blood cells, including mature blood cells, hematopoietic cells, and the LCs in lymph nodes (LNs) and lymphoid tissues (LTs) are all anoikis-resistant. Namely, a LC can survive no matter whether it is in bloodstream, tissue, or lymph. Blood cells and LCs express also adhesion molecules; however these molecules are not used for anchoring permanently to other cells. A LC can survive no matter whether these molecules are expressed or not. For example, effector LCs (immune cells) and granulocytes can express some adhesion molecules during tissue inflammation. These molecules include CD11a/CD18, CD11b/CD18, CD11c/CD18, VLA-1, VLA-4, L-selectin, CD15, CD15s, and P-selectin. However, these molecules are not expressed on the LCs that do not take part in inflammation. Thus, most cell adhesion molecules are not essential for the survival of a LC. Failure of expression of these molecules does not necessarily cause cell death of LC.

Secondly, most CD (cluster of differentiation) molecules on LCs are expressed only at certain developing stages of cells. Namely, they are expressed only when there are inducers. Some molecules expressed in LCs of earlier stage are not expressed in LCs of later stage, and vise verse. For example, CD19, MCH II, and CD20 are expressed in B-pro-lymphocytes, but they are not expressed in plasma cells (effector B-cells). In contrast, CD10 is produced in plasma cells but not produced in B-pro-lymphocytes (Rudin, 1998). This means that these CD molecules are not essential for cell survival although they are important for cell differentiation. Failure of expression of these CD molecules does not necessarily cause cell death.

Thirdly, lymphocytes are the smallest nucleated cells in our body. A mature lymphocyte, including naïve cell, memory cell, and effector T-cell, is much smaller than other types of cells including tissue cell, granulocyte, and monocyte. A mature lymphocyte has a normal-sized nucleus but little cytoplasm. This indicates that a lymphocyte expresses much fewer intracellular proteins than other types of cells. For example, a granulocyte and a monocyte have both large cytoplasms, because they produce constitutively a large number of proteins and enzymes for their functions: digesting pathogens and foreign substances. A lymphocyte is small possibly because it is a highly specialized cell. The main function of an activated lymphocyte is to produce antigen-specific immunoglobulin (Ig) in B-cells or antigen-specific TCR in T-cells. Therefore, many genes that are activated in a granulocyte/monocyte or a tissue cell may be not activated in a LC. Failure of expression of these genes may affect a granulocyte/monocyte or a tissue cell but not a LC.

In summary, a LC has three cellular characteristics different from a tissue cell: anchor-independence for cell survival, inducible expression of cell surface molecules, and expression of fewer genes than other cells as being the smallest cell (Box 2).

**Box 2: Three cellular characteristics of a LC distinct from a tissue cell**

- Anchor-independence for cell survival

- Inducible expression of cell surface molecules

- Expression of fewer genes than other cells as being the smallest cell

## 4.2   Higher survivability of a LC from DNA changes than a tissue cell

As a type of blood cell, a LC has three characteristics different from a tissue cell. These characteristics are important for us to understand childhood leukemia and lymphoma, since they are related to the rate of cell transformation of a LC. By these characteristics, a LC may have higher tolerance to DNA changes than a tissue cell. Firstly, most of the cell adhesion molecules and CD molecules on a LC may be not essential for cell survival. Failure of expression of these molecules does not necessarily lead to cell death. Namely, if a DNA change occurs to a part of a DNA that codes one of these molecules, the DNA change does not necessarily cause cell death. In another word, a LC is tolerant to the DNA changes occurred to genes of these molecules. Hence, a LC may have higher survivability from DNA changes than a tissue cell. Some forms of DNA changes that are fatal for a tissue cell may be not fatal for a LC. A LC may be able to survive from more number of DNA changes than a tissue cell. Granulocytes and monocytes are also blood cells, thus they may also have higher tolerances to DNA changes than tissue cells.

Secondly, being the smallest cell, a lymphocyte may have higher tolerance to DNA changes than a granulocyte/monocyte. A mature lymphocyte has little cytoplasm, indicating that the cell produces much fewer proteins than a granulocyte/monocyte which has large cytoplasm. This means that most genes and most parts of DNAs in a lymphocyte are "switched-off" permanently. If a DNA change occurs to a "switched-off" part of a DNA, the DNA change will be also "switched off". Thus, many PDMs and MECCs in a lymphocyte may be silent. A DNA change that has an effect on a granulocyte/monocyte may have no effect on a lymphocyte. Therefore, among all types of blood cells, LCs may have the highest survivability from DNA changes.

### 4.3 A differentiating LC may have higher tolerance to DNA changes than a non-differentiating LC

During the production of lymphocytes in marrow and during the activation of a lymphocyte in a LN/LT, some LCs are differentiating and proliferating whereas others are mature and non-proliferative. We name the LCs that are in process of differentiation as "differentiating LCs" and the LCs that are not proliferative as "non-differentiating LCs". Differentiating LCs are immature cells and they include progenitor cells, lymphoblasts, pro-lymphocytes, centroblasts, immunoblasts, and plasmablasts (Figure 1). Non-differentiating LCs are mature cells, and they include naïve lymphocytes, centrocytes, memory cells, and effector cells.

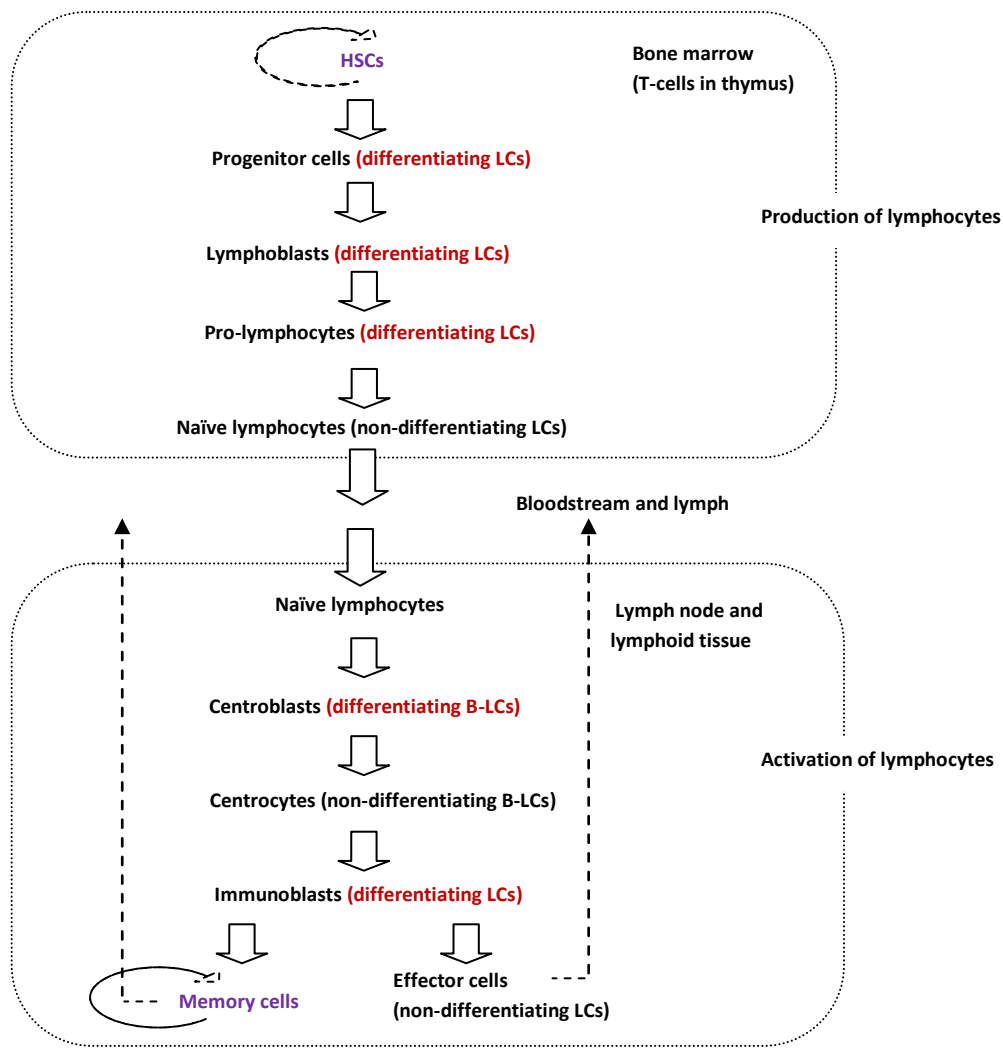

**Figure 1. Differentiating LCs and non-differentiating LCs**

During the production of lymphocytes in marrow and during the activation of a lymphocyte in a LN/LT, some LCs are differentiating and proliferating whereas others are mature and non-proliferative. In this paper, the LCs that are in process of differentiation and proliferation are called "differentiating LCs", and they include progenitor cells, lymphoblasts, pro-lymphocytes, centroblasts, immunoblast, and plasmablasts. The LCs that are mature and non-proliferative are called "non-differentiating LCs", and they include naïve lymphocytes, memory cells, effector cells, and centrocytes. Because of immaturity on cell functions, a differentiating LC may have higher tolerance to DNA changes than a non-differentiating LC. HSCs are stem cells for all LCs, but they are sometime at G0 stage. A memory cell is at G0 stage most time, but it can become proliferative when activated by an antigen. HSCs and memory cells are both mature (non-differentiating at most time) cells.

HSCs are ancestors for all LCs and myeloid cells (MCs). HSCs are stem cells, but they are sometime at **G0** stage. A memory cell is at **G0** stage most time, but it can become proliferative when activated by an antigen. HSCs and memory cells are both mature (non-differentiating at most time) cells. Distinguishing between these two groups of LCs is important, because differentiating LCs may have higher tolerance to DNA changes than non-differentiating LCs.

Firstly, a non-differentiating cell has high integrity with respect to cell functions, thus it is sensitive to any change occurred to the cell. Differently, a differentiating (immature) LC may be less alert to a change to the cell because of deficiency of cell functions. Namely, a differentiating LC may have no reaction to a DNA change, which may be fatal to a non-differentiating LC. Secondly, for achieving full functions, a mature cell expresses often more types of molecules than its precursor immature cells. Thus, more genes are normally activated in mature cells. If a DNA change occurs to a gene that is only activated in mature cells, the DNA change can only affect mature cells but not immature cells. Taken together, because of immaturity on cell functions and activation of fewer genes, a differentiating LC may have higher tolerance to a DNA change than a non-differentiating LC.

In summary, a LC may have higher survivability from DNA changes than a tissue cell by three cellular characteristics: **A.** anchor-independence for survival; **B.** inducible expression of cell surface molecules; and **C.** expression of fewer genes than other cells as being the smallest cell. In addition, due to the immaturity on cell functions, a differentiating (immature) LC may have higher tolerance to DNA changes than a non-differentiating (mature) LC. Therefore, on the tolerance to DNA changes, differentiating LC > non-differentiating LC > granulocyte/monocyte > tissue cell. We think that, it is the higher tolerance to DNA changes that makes a LC have a risk to be transformed by a chromosome change.

## V. A LC may require obtaining fewer cancerous properties for transformation than a tissue cell

Neoplasm cells can proliferate unlimitedly and even migrate to other organs because they have cancerous properties. A somatic cell needs to obtain these properties by DNA changes for a neoplasmic transformation. A cancer cell can have many cancerous properties, but the most important five are: stimulator-independent mitosis (SIM), loss of cell-contact inhibition (LCCI), production of matrix metalloproteinase (PMM), anoikis-resistance (AR), and acquired mobility (AM) (Table 2). The first two properties, namely SIM and LCCI, are essential for non-malignant transformation of a tissue cell. SIM makes a cell able to proliferate independently, and LCCI enables an unlimited cell proliferation by overcoming the cell-contact inhibition in tissue. Other properties including PMM, AR, and AM are three additional properties for a malignant tumor cell. Therefore, for non-malignant transformation into a tumor cell, a tissue cell needs to obtain two properties: SIM and LCCI. For further malignant transformation, a tumor cell needs to gain additionally one of PMM, AR, and AM (Table 2) (Wang-Michelitsch, 2015).

However, a LC has by nature some properties similar to that of a cancer cell. Firstly, as a type of blood cell, a LC is anoikis-resistant. Secondly, proliferation of developing LCs is not restricted by cell-contact inhibition. Thirdly, a LC has auto-mobility for migrating in bloodstream, lymph, and tissues. Finally, a LC can produce matrix metalloproteinases during migration (Table 2). Taken together, a LC has already by its nature four cancer-like properties: LCCL, PMM, AR, and AM. For cell transformation, a LC needs only acquire "one" more property, namely the SIM. Thus, a LC can be transformed possibly in "one step", but a tissue cell cannot. A LC can be transformed more rapidly than a tissue cell. Namely, the path to become a cancer cell is shorter for a LC than a tissue cell. In addition, cell transformation of a LC is always "malignant", because a normal LC can migrate. Therefore, in clinic, the severity of LL/lymphoma is marked not by "malignancy" but by "grade": low-grade (indolent), intermediate-grade, or high-grade (aggressive).

**Table 2. A LC may require obtaining fewer cancerous properties for cell transformation than a tissue cell**

| Cancerous properties | Properties required to obtain for cell transformation | | |
|---|---|---|---|
| | | Tissue cell | Lymphoid cell |
| Stimulator-independent mitosis (SIM) | **For non-malignant transformation** | 1 | 1 |
| Loss of cell-contact inhibition(LCCI) | | 1 | 0 |
| Production of matrix metalloproteinase (PMM) | **For further malignant transformation** | 1 | 0 |
| Anoikis-resistance (AR) | | 1 | 0 |
| Acquired Mobility (AM) | | 1 | 0 |

## VI. Transformation of a LC can be driven by accumulation of PDMs, MECCs, and IECC(s) or by a GECC

Our analyses on cell properties of LCs reveal two points: **A.** a LC may have higher survivability from DNA changes than a tissue cell; and **B.** a LC requires obtaining fewer cancerous properties for cell transformation than a tissue cell. By such a cellular nature, a LC can be transformed not only by accumulation of PDMs, MECCs, and IECC(s), but also possibly by a GECC in "one step".

## 6.1  All LCs can be transformed by accumulation of PDMs and MECCs

For a LC, acquisition of the property "SIM" is sufficient for cell transformation. A LC can acquire SIM by accumulation of PDMs/MECCs. Long-term accumulation of PDMs/MECCs proceeds mainly in HSCs and memory cells. However, all the offspring cells can inherit the DNA changes accumulated in HSCs and/or memory cells. Further mutations can take place in any one of the offspring cells. If a final PDM/MECC, together with some inherited PDMs/MECCs, makes an offspring cell obtain SIM, cell transformation may occur. For example, in FL, the final driver mutation is generated in a centrocyte in a LN/LT, but other driver DNA changes may be generated in its precursor HSCs and LCs. All LCs can be transformed by accumulation of PDMs and MECCs.

Notably, a lymphocyte has little cytoplasm, suggesting that the cell produces few intracellular proteins. In another word, in a LC, most parts of DNAs may be "switched-off". This implies that many PDMs/MECCs may have no effect on a LC. Namely, most PDMs/MECCs in a LC do not contribute to cell transformation. If a LC has neither IECC nor GECC, it may require a large number of PDMs and MECCs for cell transformation. In our view, a LC that has passed a LN/LT may have higher risk than a lymphoblast/pro-lymphocyte to be transformed by accumulation of PDMs and MECCs. Three reasons are: **A.** the LCs in a LN/LT have higher risk of DNA breaks and generations of PDMs/MECCs by pathogen-infections; and **B.** the LCs in a LN/LT are downstream cells of lymphoblasts/pro-lymphocytes, and they have more PDMs/MECCs than the LCs in marrow; and **C.** a downstream cell of memory cell can inherit PDMs/MECCs not only from HSCs but also from memory cells. CLL is a type of leukemia originated from a mature B-cell; however CLL may arise more often from a memory B-cell rather than a naïve cell. ATLL is a form of lymphoma/leukemia originated from a mature T-cell; however it may develop more often from an offspring cell of memory T-cell.

## 6.2   A differentiating LC may have a risk to be transformed by a GECC in "one step"

Clinic data show that CCs are more frequent in pediatric LLs/lymphomas than in adult cases. The difference on occurring age between BL and nasopharyngeal cancer reveals that a LC can be transformed more rapidly than an epithelial cell and this rapider transformation is more likely driven by CCs.  By affecting one or more genes, a GECC makes cell death in most cases. However, in a rare case, for a certain type of cells, if a form of GECC is not fatal but rather enables a cell to acquire the property of SIM (stimulator-independent mitosis), cell transformation may take place. Such a transformation is in "one step", because it is achieved in one generation of cell. Since a non-differentiating (mature) LC cannot survive from a GECC, rapid transformation may occur only to a differentiating (immature) LC. A NCC affects multiple genes, thus it is often a form of GECC.  For example, aneuploid (a form of NCC) may be a form of GECC, and it is responsible for the cell transformation of lymphoblast in many ALLs (Al-Bahar, 2010; Braoudaki, 2012). Some forms of SCCs, including the t(8;14) in BL and the t(2;5) in ALK[+]-ALCL, may be also forms of GECCs. They are respectively responsible for the rapider transformations in these two forms of childhood lymphomas (Weniger, 2006).

### 6.3 An IECC may accelerate cell transformation driven by accumulation of PDMs and MECCs

An IECC is different from a GECC, because it cannot alone cause cell death or cell transformation. However, with stronger effect than a PDM/MECC, an IECC can accelerate cell transformation driven by accumulation of PDMs and MECCs. An IECC accelerates cell transformation probably by affecting an oncogene or a tumor suppressor gene. This accelerated transformation occurs mainly to a differentiating LC, because a non-differentiating (mature) LC is not tolerant to an IECC. In our view, Ph translocation is a form of IECC. Ph translocation may be the accelerating factor that enables CML to occur at any age. In CML development, Ph translocation contributes to cell transformation by generating fusion gene of *BCR-ABL1* and activating ABL1 permanently. ABL1 is a tyrosine kinase that can promote cell proliferation. *ABL1* is a type of oncogene.

### VII.   Our hypothesis: a LC may have three pathways on cell transformation

Due to some cellular characteristics, a LC may have higher survivability from DNA changes and require obtaining fewer cancerous properties for cell transformation than a tissue cell. Thus, a LC can be transformed not only by accumulation of PDMs, MECCs, and IECC(s), but also possibly by a GECC. On this basis, we establish the following hypothesis: *a LC may have three pathways on cell transformation: a slow, a rapid, and an accelerated* (Box 3). Slow pathway is driven by accumulation of PDMs and MECCs though many generations of cells. Rapid pathway is driven by a GECC in "one step", namely in one generation of cell. Accelerated pathway is driven by accumulation of PDMs, MECCs, and IECC(s) through a few generations of cells. Distinguishing between these three pathways of cell transformation of a LC is important in understanding LL/lymphoma, because it helps answering two questions: **A.** why some forms of LLs/lymphomas occur in young children; and **B.** why pediatric forms of LLs/lymphomas are often aggressive.

**Box 3. Our hypothesis: a LC may have three pathways on cell transformation**

---

- **Slow pathway**: by accumulation of PDMs and MECCs through many generations of cells

- **Rapid pathway**: by a GECC in one step in one generation of cell

- **Accelerated pathway**: by accumulation of PDMs, MECCs, and IECC(s) through a few generations of cells

PDM: point DNA mutation                      GECC: great-effect chromosome change
MECC: mild-effect chromosome change          IECC: intermediate-effect chromosome change

---

## 7.1 Slow pathway: by accumulation of PDMs and MECCs through many generations of cells

Similar to a tissue cell, a LC can obtain the property of stimulator-independent mitosis (SIM) and undergo cell transformation by accumulation of PDMs and MECCs. Although the final PDM/MECC occurs in the first transformed cell, most of other driver PDMs/MECCs are generated in precursor cells including HSCs and/or memory cells. Long-term accumulation of PDMs and MECCs takes place mainly in regenerable HSCs and memory cells, as a result of repeated cell injuries and repeated cell proliferation. Thus, accumulation of PDMs and MECCs needs to proceed over many generations of HSCs and memory cells, and it is paid by death of a great deal of injured cells (Wang-Michelitsch, 2015). Therefore, the transformation driven by accumulation of PDMs and MECCs is a slow process. We name this pathway of transformation as "slow pathway". A cell transformation via this pathway takes place mainly in old people and rarely in children (Figure 2). As an effect of accumulation of damage and DNA changes, the incidence of cell transformation via this pathway increases with age (Figure 2 and Figure 3). All LCs can be transformed via this pathway. Non-differentiating LCs can be transformed mainly via this pathway. A tissue cell can only be transformed via slow pathway.

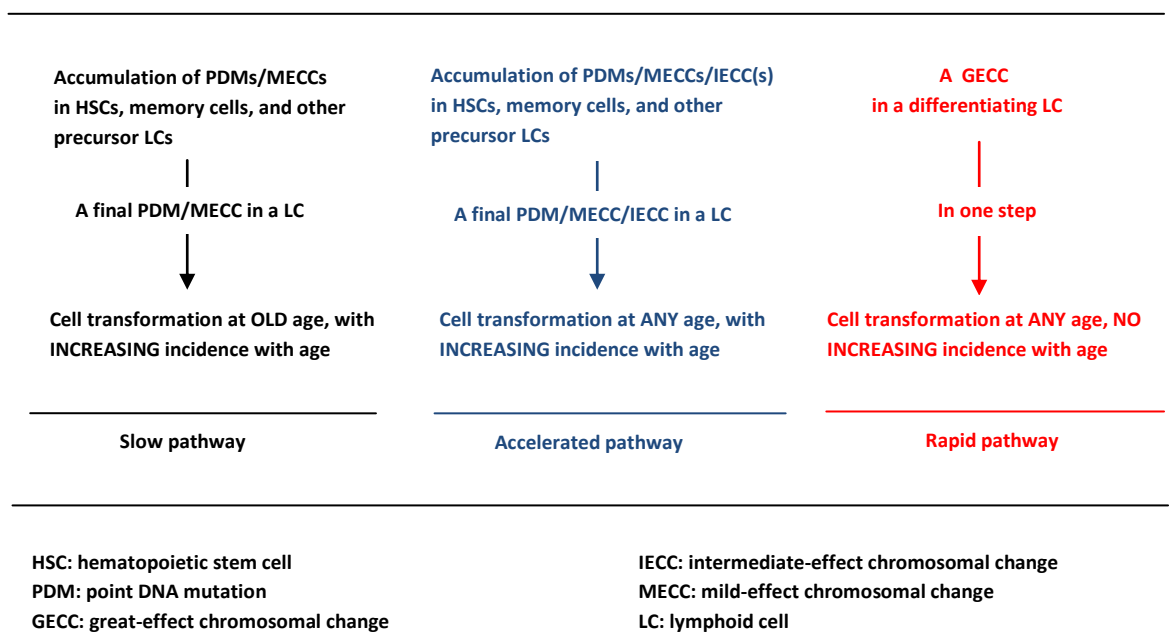

HSC: hematopoietic stem cell
PDM: point DNA mutation
GECC: great-effect chromosomal change

IECC: intermediate-effect chromosomal change
MECC: mild-effect chromosomal change
LC: lymphoid cell

**Figure 2. Hypothesized three pathways of a LC on cell transformation**

Due to some cellular characteristics, a LC may have higher survivability from DNA changes and require obtaining fewer cancerous properties for cell transformation than a tissue cell. Thus, a LC can be transformed not only by accumulation of PDMs/MECCs/IECC(s), but also possibly by a GECC. On this basis, we hypothesize that: a LC may have three pathways on cell transformation: a **rapid pathway** by a GECC, a **slow pathway** by accumulation of PDMs and MECCs, and an **accelerated pathway** by accumulation of PDMs, MECCs, and IECC(s). In slow and accelerated pathway, accumulation of PDMs, MECCs, and IECC(s) takes place mainly in HSCs and memory cells. But the final PDM/MECC/IECC takes place in the first transformed cell. Cell transformations via different pathways occur at different ages. A transformation via slow pathway occurs mainly in adults and has increasing incidence with age. A transformation via rapid pathway can take place at any age and has no increasing incidence with age. A transformation via accelerated pathway can take place also at any age, but it has increasing incidence with age.

Adult forms of LLs/lymphomas, including CLL, FL, MALTL, MCL, and ATLL, occur mainly in old people, and they have increasing incidences with age. Thus, they develop mainly via slow pathway. The DLBCL in adults may also develop via this pathway. The LLs/lymphomas that have a peak incidence in children or young people, such as ALL, BL, T-LBL, ALK[+]-ALCL, and HL, may not develop via this pathway.

For a LL/lymphoma that develops by accumulation of PDMs and MECCs, the PDMs/MECCs in different cases can be quite different. Even in one patient, different tumor cells may have different PDMs/MECCs. The heterogeneity of tumor cells on DNA changes may be a main cause for the treatment-resistance of adult cancers. In addition, in an adult form of LL/lymphoma, the "sister cell" and some of "cousin cells" of the first transformed cell have similar DNA changes to this transformed cell, because they all have the same precursor HSCs and LCs. Thus, these cells are a kind of pre-cancer cells. A pre-cancer cell can be promoted to transform by chemotherapy or radiotherapy (Rothkamm 2002). Hence, existence of pre-cancer cells may be a cause for the relapse of adult cancers after treatment.

## 7.2 Rapid pathway: by a GECC in "one step" in one generation of cell

A LC requires only obtaining one more property, namely the "SIM", for transformation. A differentiating (immature) LC may be tolerant to a GECC due to its higher survivability from DNA changes than other cells. Thus, a differentiating LC may have a risk to be transformed directly by a GECC. Such a cell transformation can take place in "one step", namely in one generation of cell. Thus, the transforming pathway by a GECC is a rapid pathway (Figure 2). The cell transformation via this pathway can take place at any age, including young children, adolescents, and adults. Such a rapid transformation does not need accumulation of DNA changes. Thus, the incidence rate does not increase with age, but rather peaks at certain age(s) (Figure 2 and Figure 3). The driver GECC in rapid transformation is generated in the first transformed cell. There may be also PDMs/MECCs in this transformed cell, but generation of the GECC is not related to these PDMs/MECCs. The PDMs/MECCs do not necessarily contribute to cell transformation driven by a GECC. With lower tolerance to DNA changes, a non-differentiating (mature) LC cannot be transformed by a GECC via rapid pathway. Since a GECC can affect multiple genes and disturb cell differentiation, rapid transformation is often at high-grade and results in occurrence of an aggressive form of LL/lymphoma.

Some pediatric forms of LLs/lymphomas, including ALL, HL, BL, T-LBL, and ALK[+]-ALCL, can occur at any age and have no increasing incidence with age. Thus, they develop more likely via this pathway. Those forms of LLs/lymphomas that occur mainly in adults, such as CLL, FL MALTL, MCL, and ATLL, may not develop via this pathway. DLBCL occurs also in children, but DLBCL has increasing incidence with age. Thus DLBCL may not develop via this pathway. In a LL/lymphoma developed via rapid pathway, the neoplasm cells look rather homogenous on morphology. One reason is: the properties of tumor cells are mostly determined by the strong GECC, even if the tumor cells have also other PDMs/MECCs. This can explain why pediatric LLs/lymphomas have often good responses to a treatment even if they are aggressive.

### 7.3 Accelerated pathway: by accumulation of PDMs, MECCs, and IECC(s) through a few generations of cells

Cell transformation of LC can be also a co-effect of an IECC and PDMs/MECCs. With stronger impact than a PDM/MECC, an IECC can accelerate cell transformation driven by accumulation of PDMs and MECCs. Namely, the transformation driven by an IECC and PDMs/MECCs is more rapid than that in slow pathway. We name this pathway as "accelerated pathway" (Figure 2). A transformation via accelerated pathway needs to be achieved also through some generations of cells, but fewer than that in slow pathway. As discussed earlier, an IECC can occur in a LC at any time, but it has increasing occurrence in LCs with age of our body. Thus, a transformation via this pathway can take place at any age, but the incidence rate will increase with age (Figure 2 and Figure 3). A differentiating (immature) LC may be tolerant to an IECC, thus it has a risk to be transformed via this pathway. An IECC can be generated in the first transformed cell or in one of its precursor cells.

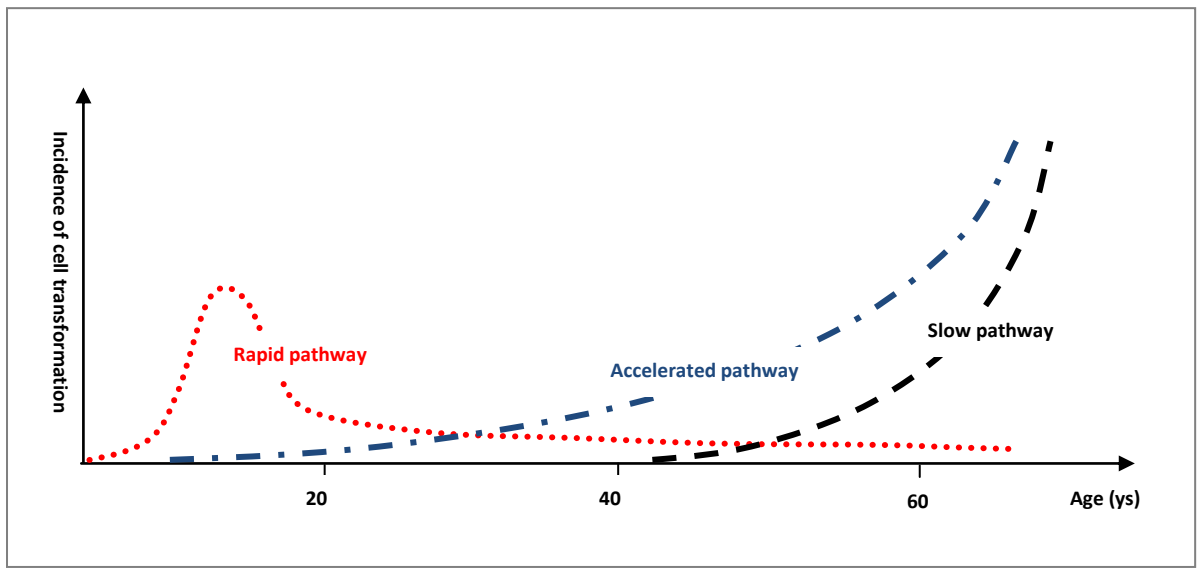

**Figure 3. Transformations of a LC via different pathways occur at different ages (a schematic graph)**

Distinguishing between three pathways of cell transformation of a LC can help us understand the age-specificity in LL/lymphoma development. The age of occurrence of cell transformation is related to the transforming pathway of a LC. A transformation via slow pathway takes place mainly in adults and has gradual increasing incidence with age (**Black line**). A transformation via rapid pathway can take place at any age and has no increasing incidence with age, but has a peak at certain age (**Red line**). A transformation via accelerated pathway can take place also at any age, but it has increasing incidence with age (**Blue line**). A differentiating (immature) LC can be transformed via all three pathways, but a non-differentiating (mature) LC is transformed mainly via slow pathway.

Ph$^+$-ALL and Ph-like ALL occur mainly in adults and older children, and their incidence rates increase with age. Thus these two subtypes of ALL may develop via accelerated pathway. Ph translocation t(9;22) and CRLF2 rearrangement may be two forms of IECCs that can accelerate cell transformation. Similarly, occurred mostly in adults but also in older children, some DLBCLs including pediatric cases may also develop via this pathway. The MYC-translocations in DLBCL cells may be IECCs. Rare in children and young people, CLL, MALTL, MCL, and ATLL may not develop via this pathway. Having no increasing incidence with age, BL, T-LBL, HL, and ALK$^+$-ALCL do not develop via this pathway.

Taken together, distinguishing between three pathways of cell transformation of a LC can help us understand the age-specificity in LL/lymphoma development. The age of occurrence of cell transformation is related to the transforming pathway of a LC. A transformation via slow pathway takes place mainly in adults and has increasing incidence with age (Figure 3). A transformation via rapid pathway can take place at any age and has no increasing incidence with age. A transformation via accelerated pathway can take place also at any age, but it has increasing incidence with age. A differentiating (immature) LC can be transformed via all three pathways, but a non-differentiating (mature) LC is transformed mainly via slow pathway. Notably, in the cell transformation via slow or accelerated pathway, the driver DNA changes are generated some in precursor HSCs/LCs in marrow and some in precursor LCs/memory cells in LNs/LTs. Thus, the cell injuries of HSCs/LCs in marrow and that of LCs in LNs/LTs may be all associated with the development of an adult LL/lymphoma.

## VIII. Three grades of transformation of a LC: low, high, and intermediate

In clinic, the severity of a LL/lymphoma is marked by the grade of neoplasm: low-grade, high-grade, or intermediate-grade. The grade of a LL/lymphoma is related to the grade of cell transformation of LC. A LC may have three grades on transformation: low-grade, high-grade, and intermediate-grade. For a differentiating LC, cell transformation can be differentiation-affected or differentiation-not-affected. If cell differentiation is not at all affected, the transformation is at **low-grade**. If cell differentiation is severely affected, the transformation is at **high-grade.** There is still a third possibility: if cell differentiation is partially affected, the transformation is at **intermediate-grade.** Due to higher tolerance to DNA changes, a differentiating LC may have a risk to be transformed directly at high-grade or intermediate-grade. However, for a non-differentiating LC, cell transformation begins often at low-grade as that seen in CLL and FL. Similarly, a tissue cell including stem cell and progenitor cell has to be transformed firstly at low-grade, then to high-grade.

For a tumor developed from a tissue cell, the grade of transformation determines whether the tumor is non-malignant (low-grade) or malignant (intermediate-grade or high-grade). Differently, the grade of transformation of a LC determines not only the severity but also the form or subtype of a LL/lymphoma. For example, transformation of a lymphoblast at low-grade may result in occurrence of lymphoid proliferative disorder (LPD), but that at intermediate-grade or high-grade may result in occurrence of acute lymphoblastic leukemia (ALL) (Table 3). Transformation of a centroblast at low-grade may lead to FL development but that at intermediate-grade or high-grade may result in BL development. Transformation of a B-immunoblast at low-grade may lead to occurrence of large B-cell lymphoma (LBCL), but that at intermediate-grade or high-grade may result in DLBCL development. Non-differentiating LCs can be transformed mainly via slow pathway and at low-grade, thus chronic and indolent forms of LLs/lymphomas, such as CLL, MALTL, MCL, and ATLL, do not occur in children.

A transformation via slow pathway and accelerated pathway begins often by low-grade, in which the cell differentiation of tumor cells is not affected. For example, low-grade transformation of a lymphoblast may firstly result in development of LPD. However, further high-grade transformation of one of LPD cells by Ph translocation (t(9;22)) may lead to development of Ph$^+$-ALL (Table 3). Likely, transformation of a HSC at low-grade may result in clonal hematopoiesis; however further intermediate- or high-grade transformation of a clonal HSC may result in occurrence of myelodysplastic syndrome (MDS). *In situ* follicular neoplasia (ISFN) and *in situ* mental cell neoplasia (ISMCN) are two examples of pre-lymphomas. They are early events of FL and MCL respectively. These neoplasia changes should be results of clonal proliferation of LCs by low-grade cell transformation.

**Table 3. The grades of cell transformation of LCs in different forms of LLs/lymphomas**

| Cell of origin | Form of LL/lymphoma | |
|---|---|---|
| | At low-grade transformation | At intermediate-grade or high-grade transformation |
| HSC | Clonal hematopoiesis | MDS? |
| Lymphoblast | Lymphoid proliferative disorder | ALL |
| Naïve lymphocyte | MCL | ------ Cell death |
| Centroblast | FL? | BL |
| Immunoblast | Large B-cell lymphoma (LBCL) ? | DLBCL |
| Memory B-cell | CLL | ------ Cell death |

## IX.  Summary

We have discussed in the present paper the causes and the mechanism of cell transformation of a LC in development of lymphoid leukemia (LL) and lymphoma. Here we give an outline of our work on studying LL/lymphoma (Figure 4). Four papers on this work are being prepared, and they are paper A, B, C, and D. The present paper is paper B.

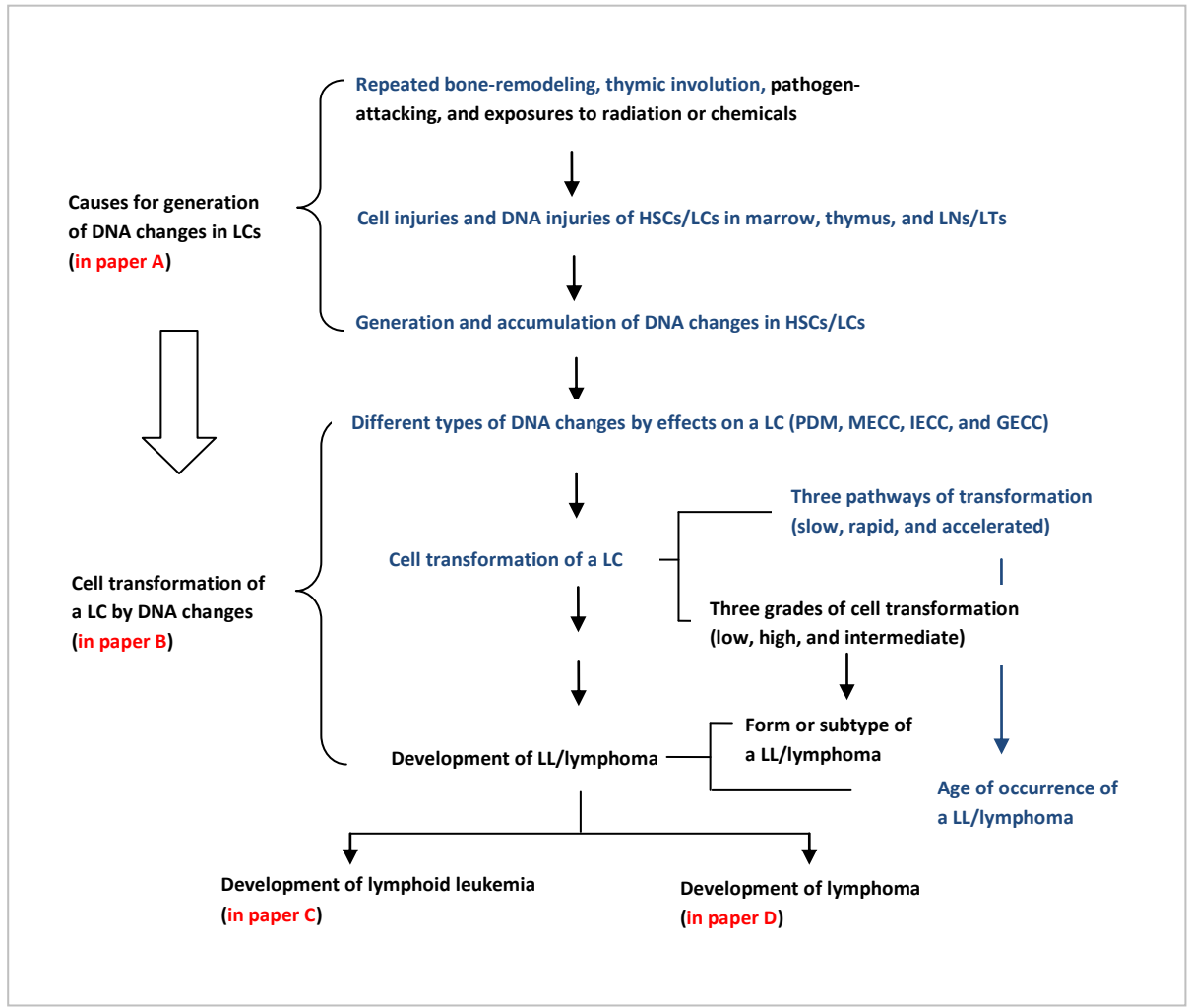

**Figure 4. Our understanding of the pathogenesis of lymphoid leukemia and lymphoma**

This is the outline of our work on studying the pathogenesis of LL/lymphoma. DNA changes are the trigger for cell transformation, and they are generated as consequences of cell injuries and DNA injuries. Apart from pathogen-infections and exposures to radiation or chemicals, repeated bone-remodeling during bone-growth and thymic involution may be two other potential sources of cell injuries of HSCs and LCs in marrow and in thymus. Generation and accumulation of DNA changes in HSCs and LCs are results of repeated cell injuries and repeated cell proliferation.

DNA changes can be classified into four groups by their effects on a cell: PDM, MECC, IECC, and GECC. By different types of DNA changes, a LC can be transformed at different grades (low, high, or intermediate) and via different pathways (slow, rapid, or accelerated). The grade of cell transformation determines the form or subtype of a LL/lymphoma. The pathway of cell transformation determines the age of occurrence of a LL/lymphoma. **The parts in blue color are our hypotheses.** Four papers on this work (**paper A, B, C,** and **D**) are currently under preparation. The present paper is **paper B.**

## X. Conclusions

We discussed in this paper the effects of different types of DNA changes on a LC and the cellular characteristic of LCs. A LC is different from a tissue cell by three characteristics: anchor-independence on survival, inducible expression of cell surface molecules, expression of fewer genes as being the smallest cell. Due to these characteristics, a LC may have higher survivability from DNA changes and require obtaining fewer cancerous properties for cell transformation than a tissue cell. Thus, a LC can be transformed not only by accumulation of PDMs, MECCs, and IECC(s) but also possibly by a GECC. On this basis, we hypothesize that a LC may have three pathways on transformation: a slow, a rapid, and an accelerated. The age of occurrence of LL/lymphoma may be determined by the transforming pathway of a LC. Pediatric forms of LLs/lymphomas may develop via rapid or accelerated pathway, and adult forms may develop via slow or accelerated pathway.

To verify our hypotheses on the higher survivability of differentiating LCs than tissue cells and the rapider pathway of cell transformation of a LC, experimental researches can be undertaken to study the effects of different types of DNA changes on proliferative LCs. Experiments can be done by using cultured lymphocytes and tissue cells (such as epithelia cells). DNA injuries of cultured cells can be made by UV-radiations.

It is a tragedy when a child develops LL or lymphoma. Our analysis shows that this tragedy may be partially related to the nature of lymphocytes different from tissue cells. Lymphocytes are important for our immunity as highly-treated "soldiers". However, the higher tolerance of differentiating LCs to DNA changes than tissue cells may make them more susceptible to cell transformation.